\documentclass[a4paper]{jpconf}
\usepackage{graphicx}
\usepackage{amsmath,amsfonts,amssymb}
\usepackage{marvosym}
\usepackage{txfonts}
\usepackage[T1]{fontenc}
\usepackage{xcolor}
\usepackage{mathtools}

\newcommand{\dd}{\partial}
\newcommand{\ff}{\mathcal{F}}
\newcommand{\ot}{\otimes}

\newcommand{\uf}{\mathcal{U}^\ff(\mathfrak{g})}
\newcommand{\mm}{\mathcal{M}}
\newcommand{\oo}{\mathring}
\newcommand{\lie}{\mathcal{L}}

\newcommand{\vp}{\varphi}

\begin{document}


\title{Towards gravitational QNM spectrum from quantum spacetime}


\author{Nikola Herceg$^1$, Tajron Juri\'{c}$^2$, Andjelo Samsarov$^3$ and Ivica Smoli\'{c}$^4$}

\address{$^{1,2,3}$ Rudjer Bo\v{s}kovi\'c Institute, Bijeni\v cka  c.54, 10000 Zagreb, Croatia}

\address{$^4$ Department of Physics, Faculty of Science, University of Zagreb, 10000 Zagreb, Croatia}

\ead{$^1$nherceg@irb.hr, $^2$tjuric@irb.hr, $^3$asamsarov@irb.hr, $^4$ismolic@phy.hr} 


\begin{abstract}
The effective potential for the axial mode of gravitational wave on noncommutative Schwarzschild background is presented.
	Noncommutativity is introduced via deformed Hopf algebra of diffeomorphisms by means of a semi-Killing Drinfeld twist. 
	The analysis is  performed up to the first order in perturbation of the metric and noncommutativity parameter.
	This results in a modified Regge-Wheeler potential with   the strongest differences in comparison to the classical Regge-Wheeler potential being near the horizon.
\end{abstract}

\section{Introduction}

One of the greatest puzzles of modern physics revolves around reconciliation of two highly successful theories - quantum mechanics and general relativity. Many different approaches have been launched to this day in order to confront this awe-inspiring but at the same time incredibly challenging task. They include string theory, loop quantum gravity, noncommutative (NC) gravity, causal set theory and others.
Most of them propose that the classical notion of spacetime needs to be modified at small scales,  leading to a concept of quantum spacetime.
Moreover,  they are   associated with  a notion of Lorentz symmetry breaking which appears to crop up in their  low-energy limit as a remnant of quantum gravity. 
The Lorentz invariance violation therefore presents one of the most striking signals of quantum structure of spacetime and indirectly of quantum gravity.

It therefore doesn't come as surprise
 that the search  for a quantum nature of spacetime included proposals
for novel physical phenomena arising due to  the effects of quantum gravity  in the low-energy regime involving  Lorentz violation as a central part.
A suitable framework for carrying out this kind of study  has either  been provided  by
the Lorentz Invariance Violation (LIV) models \cite{liv},\cite{liv1}, or by the  Doubly Special Relativity (DSR) theories which preserve
the relativity principle, but implement it by means of deformed Lorentz transformations \cite{dsr},\cite{dsr1}. 

Over a past few decades four major channels of cosmic messengers have been crystallized as  suitable avenues for testing and identifying the quantum nature of spacetime.
They include cosmic neutrinos, ultra high energy cosmic rays, gamma ray bursts and gravitational waves \cite{QGphen}.
It is the latter channel that we shall focus on in the present work.
In particular, we shall outline the main steps in constructing  NC theory of gravity in a bottom-up approach by utilising
the framework of Hopf algebra deformation by means of a Drinfeld twist \cite{Aschieri:2005zs},\cite{Aschieri:2005yw},\cite{Aschieri:2009qh},\cite{schenkel}.
The construction will be applied to the linearized gravitation perturbation theory of the Schwarzschild black hole background, thus setting the ground  for 
determining  the corrections in the black hole QNM spectrum due to quantum spacetime.

\section{Hopf algebra deformation by twist and NC differential geometry} 

Let us assume that the universal enveloping algebra ${\mathcal{U}}(\mathfrak{g}) $ of the Lie algebra $\mathfrak{g}$ of diffeomorphisms may be organized into a Hopf algebra,
an algebraic structure that besides the multiplication and unital maps also involves additional structural maps like coproduct $\Delta$, antipode $S$ and counit 
$\epsilon $, defined for each element of the Hopf algebra.
We further assume that there is an invertible element $\ff$  in  ${\mathcal{U}}(\mathfrak{g}) \otimes   {\mathcal{U}}(\mathfrak{g}).$ This element that we call the twist 
and its inverse may respectively be  written as 
 ${\mathcal{F}} = f^{\alpha} \otimes f_{\alpha},$  and  
   $  {\mathcal{F}}^{-1} = {\bar{f}}^{\alpha} \otimes {\bar{f}}_{\alpha},$  with the summation over $\alpha$ being understood. The twist element may be used to deform the initial Hopf algebra  ${\mathcal{U}}(\mathfrak{g}) $  by means of the  transformations
\begin{equation}    \label{twistdeformation}
	\Delta^\ff(h) = \ff \Delta(h) \ff^{-1}, \quad S^\ff(h) = \chi S(h) \chi^{-1}, \quad
	\epsilon^\ff(h) =  \epsilon(h),    
\end{equation}
where $\chi = S({\bar{f}}^{\alpha})  {\bar{f}}_{\alpha}$.
In this way the  Hopf algebra ${\mathcal{U}}(\mathfrak{g}) $ goes into a new, deformed structure that may be denoted as $\uf,$ and that will also be a Hopf algebra if (by adopting the Sweedler's notation) the twist  element satisfies
\begin{eqnarray}   \nonumber
     f^{\beta} f^{\alpha}_{(1)} \otimes f_{\beta} f^{\alpha}_{(2)} \otimes f_{\alpha} &=& f^{\alpha}  \otimes f^{\beta} f_{\alpha (1)} \otimes f_{\beta} f_{\alpha (2)}, \nonumber \\
    {\bar{f}}^{\alpha}_{(1)} {\bar{f}}^{\beta} \otimes  {\bar{f}}^{\alpha}_{(2)} {\bar{f}}_{\beta} \otimes {\bar{f}}_{\alpha} &=& 
  {\bar{f}}^{\alpha}  \otimes {\bar{f}}_{\alpha (1)}   {\bar{f}}^{\beta}  \otimes {\bar{f}}_{\alpha (2)}  {\bar{f}}_{\beta},   \label{2cocyclecondition}
\end{eqnarray}
which is known as the $2$-cocycle condition. The twist satisfying these conditions is called a Drinfeld twist. The deformation  (\ref{twistdeformation}) induces also a deformation at the level of the module algebra of ${\mathcal{U}}(\mathfrak{g}),$ which is $\mathcal{A} = C^\infty(\mm).$
In particular, the multiplication map is deformed from a pointwise multiplication to a $\star$-multiplication,
\begin{equation} \label{starproductdef}
	f \star g = \mu \circ \ff^{-1} (f \ot g) = {\bar{f}}^{\alpha}(f)  {\bar{f}}_{\alpha}(g),
\end{equation}
 for any two functions $f$ and $g$ in the algebra $C^\infty(\mm).$


Another   important concept is that of the quasitriangular Hopf algebra and the $R$-matrix,  $ R =R^{\alpha} \otimes R_{\alpha} \in \uf \otimes \uf, ~       R^{-1} = {\bar{R}}^{\alpha} \otimes \bar{R}_{\alpha} $. The $R$-matrix, besides satisfying the quasitriangularity conditions 
\begin{eqnarray}   \nonumber
      R^{\alpha}_{(1)} \otimes  R^{\alpha}_{(2)} \otimes R_{\alpha} &=& R^{\alpha}  \otimes R^{\beta}  \otimes R_{\alpha} R_{\beta}, \nonumber \\
    {\bar{R}}^{\alpha}_{(1)}  \otimes  {\bar{R}}^{\alpha}_{(2)}  \otimes {\bar{R}}_{\alpha} &=& 
  {\bar{R}}^{\alpha}  \otimes {\bar{R}}^{\beta}    \otimes {\bar{R}}_{\beta}  {\bar{R}}_{\alpha},  \label{quasitriangularity}
\end{eqnarray}
may also be related to the Drinfeld twist  ${\mathcal{F}}$ as $  R= {\mathcal{F}}_{op} {\mathcal{F}}^{-1},$ where
 $  {\mathcal{F}}_{op}    = \sigma \circ {\mathcal{F}}$  and $\sigma$ is the exchange operator defined as $\sigma(a \ot b) = b \ot a$.


For the details of the NC gravity construction, we refer the reader to \cite{Aschieri:2009qh, schenkel,glavniclanak}. 
In the rest of this section we briefly introduce the most important notions. To begin with, let ${\mathcal{A}}_{\star},    {\mathcal{\chi}}_{\star}, \Omega_{\star}, {\mathcal T}^{(p,q)}_{\star}$ respectively denote
the vector spaces of NC functions, vector fields, $1$-forms and general tensor fields.
The  $\star$-pairing, ${\langle ~, ~\rangle}_{\star} : {\mathcal{\chi}}_{\star}   \times   {\Omega}_{\star}  \rightarrow  {\mathcal{A}}_{\star}$ of vector fields and one-forms  is then introduced by
 \begin{equation}   \label{starpairing}
    {\langle u , \omega \rangle}_{\star} = 
  \langle {\bar{f}}^{\alpha}(u) , {\bar{f}}_{\alpha}(\omega) \rangle.
\end{equation}
Curiously, it appears that
  conditions of cocyclicity and quasitriangularity both give rise to the following properties of
the $\star$-pairing:
\begin{eqnarray}  \nonumber
    {\langle h \star u, \omega \star  \tilde{h} \rangle}_{\star} &=& h \star  {\langle  u, \omega  \rangle}_{\star} \star  \tilde{h},    \quad \quad  h, \tilde{h} \in  {\mathcal A}_{\star}, ~ u \in {\mathcal \chi}_{\star},  ~  \omega \in  \Omega_{\star},\\
   {\langle  u,  h \star \omega  \rangle}_{\star} &=&  {\bar{R}}^{\alpha}(h)                     \star  {\langle {\bar{R}}_{\alpha}(u), \omega  \rangle}_{\star}  \label{starpairingprop}, \\
    {\langle  \omega \otimes_{\star} u,  \tau  \rangle}_{\star} &=&    {\langle  \omega,   {\langle  u,  \tau  \rangle}_{\star}  \rangle}_{\star}  
   \quad \quad    u \in {\mathcal \chi}_{\star},  ~  \omega \in  {\mathcal T}^{(0,p)}_{\star}, ~ \tau \in  {\mathcal T}^{(q,s)}_{\star},  \quad (q > p).  \nonumber
\end{eqnarray}
NC connection  is introduced as a linear map,   $ \nabla^{\star}:  {\mathcal{\chi}}_{\star}   \longrightarrow  \Omega_{\star}   \otimes_{\star}  {\mathcal{\chi}}_{\star},  $  satisfying the 
standard Leibniz rule, 
$  {\nabla}^{\star} (h \star v) = dh \otimes_{\star} v + h \star {\nabla}^{\star} v,$
for all  $h \in {\mathcal{A}}_{\star}$  and  $v \in {\mathcal{\chi}}_{\star}.$ 
The notion of the covariant derivative is closely related to that of the connection. In particular,
 a covariant derivative $\nabla^{\star}_u$  along the vector field $u,$ for any  $u \in {\mathcal{\chi}}_{\star},$ 
may be defined as
\begin{equation}  \label{nccovariantderivative}
   \nabla^{\star}_u v = {\langle  u,  \nabla^{\star} v  \rangle}_{\star},    \quad  \mbox{for all $v \in {\mathcal{\chi}}_{\star}.$}
\end{equation}
  Interestingly, the set of axioms that is usually required for the covariant derivative,
\begin{equation}   \label{covderaxioms}
\begin{split}
  \nabla^{\star}_{u+v} z &= \nabla^{\star}_{u} z + \nabla^{\star}_{v} z,    \\
 \nabla^{\star}_{g \star u} v &= g \star \nabla^{\star}_{u} v,     \\
 \nabla^{\star}_{u} (g \star v) &= {\mathcal L}^{\star}_u (g ) \star v + {\bar{R}}^{c} (g) \star    
   {\nabla}^{\star}_{\bar{R}_{c} (u)} v,   
\end{split}
\end{equation}
for all $~ u, v, z \in {\mathcal{\chi}}_{\star} ~$
and $~ g \in {\mathcal{A}}_{\star},$ does not need to be postulated here, since it follows naturally  as a mere consequence of the $\star$-pairing relations (\ref{starpairingprop}). Here ${\mathcal L}^\star_u (g) = {\mathcal L}_{\bar{f}^\alpha(u)} \big(\bar{f}_\alpha (g)\big).$


Consider now  a $\star$-dual basis, ${\langle \dd_\mu , dx^\nu \rangle}_{\star} = \delta_\mu{ }^\nu,$ consisting  of  two mutually dual local frames,
that of vector fields $\{ \partial_ {\mu}\}$ and that of $1$-forms $\{ d x^{\nu} \}.$   The coefficients of affine connection $\Gamma^{\star \mu}_{~~~\nu \lambda}$ in noncommutative theory  are then consistently  given as
\begin{equation}  \label{afcoefv}
   \nabla^{\star}_{\partial_\mu} \partial_{\nu} \equiv \nabla^{\star}_{\mu} \partial_{\nu} = \Gamma^{\star   \lambda}_{~~~\mu \nu} \star \partial_{\lambda}, \quad
   \nabla^{\star}_{\partial_\mu} dx^{\nu} \equiv \nabla^{\star}_{\mu} dx^{\nu} =  - \Gamma^{\star  \nu}_{~~~\mu \lambda} \star dx^{\lambda}.
\end{equation}
Other crucial  concepts such as  the NC torsion, Riemann curvature and Ricci tensors are then straightforwardly introduced for all $~ u, v, z \in {\mathcal{\chi}}_{\star},$
\begin{equation}  \label{torsion-riemann}
\begin{split}
  T^{\star}(u,v) &=  \nabla^{\star}_{u} v - {\nabla}^{\star}_{\bar{R}^{c} (v)} {\bar{R}_{c} (u)}
  -  [u, v]_{\star}   \equiv {\langle  u \otimes_{\star} v,  T^{\star}   \rangle}_{\star},  \nonumber \\
 R^{\star}(u,v,z) &=  \nabla^{\star}_{u}\nabla^{\star}_{v} z
  -  {\nabla}^{\star}_{\bar{R}^{c} (v)}{\nabla}^{\star}_{\bar{R}_{c} (u)} z -
  {\nabla}^{\star}_{[u, v]_{\star}} z   \equiv {\langle  u \otimes_{\star} v \otimes_{\star} z,  R^{\star}   \rangle}_{\star}.  \nonumber  \\
   Ric^{\star}(u,v) &= \langle dx^\alpha, R^{\star}(\dd_\alpha, u, v) \rangle_\star.
\end{split}
\end{equation}
 and the components  $T^{\star \lambda}_{~~~\mu \nu}$ 
  and  $R^{\star ~~~~\lambda}_{~\mu \nu \sigma}$ are given by
\begin{equation}   \label{torsion-riemann-coef}
\begin{split}
  T^{\star \lambda}_{~~\mu \nu} = {\langle dx^{\lambda}, T^{\star}(\partial_{\mu}, \partial_{\nu}) \rangle}_{\star},  \quad
   R^{\star ~~~~\lambda}_{~\mu \nu \sigma} =  {\langle dx^{\lambda}, R^{\star}(\partial_{\mu}, \partial_{\nu}, \partial_{\sigma}) \rangle}_{\star}
\end{split}
\end{equation}
It can be checked that the Riemann  and torsion tensors  are $R$-antisymmetric in the first two   slots:
\begin{equation*}
	R^{\star}(u,v,z) = -R^{\star}(\bar{R}^c (v),\: \bar{R}_c (u),\: z), \qquad T^{\star}(u,v) = -T^{\star}(\bar{R}^c(v), \:\bar{R}_c (u)).
\end{equation*}
However, NC Riemann tensor  is not  R-antisymmetric in the second and the third slots, implying that NC Ricci tensor is neither $R$-symmetric nor symmetric.


Finally, the metric  $g$  is an element of   $ \Omega_{\star}   \otimes_{\star}  \Omega_{\star} $  and may be written as
\begin{equation}  \nonumber
 g= g_{\mu \nu} \star dx^{\mu} \otimes_{\star} dx^{\nu} = g^a \otimes_{\star} g_a  \in  \Omega_{\star}   \otimes_{\star}  \Omega_{\star},
\end{equation}
where the sum over $a$ is understood.
Analogously, the inverse metric $g^{-1} \equiv g^{\star} $ is an element of  ${\mathcal{\chi}}_{\star} \otimes_{\star}  {\mathcal{\chi}}_{\star}, $
$g^{-1} = g^{-1 ~ b} \otimes g_{b}^{-1 } \in   {\mathcal{\chi}}_{\star} \otimes_{\star}  {\mathcal{\chi}}_{\star} $.
These two are related by
\begin{equation}      \label{metricinversecondition}
\begin{split}
  \langle  \langle v, g \rangle_\star, g^{-1} \rangle_{\star} = \langle v, g^a \rangle_{\star} \star  \langle  g_a, g^{-1 ~ b} \rangle_{\star} \star  g_{b}^{-1 }  = v,      \quad  \quad  \mbox{for all}  \quad   v \in    {\mathcal{\chi}}_{\star}        \\
  \langle  \langle \omega, g^{-1} \rangle_\star, g \rangle_{\star} =
    \langle \omega,  g^{-1 ~ b} \rangle_\star \star  \langle  g_{b}^{-1 }, g^a \rangle_{\star} \star g_a    = \omega,
      \quad  \quad  \mbox{for all}  \quad  \omega \in \Omega_{\star}  
\end{split}
\end{equation}
The fact that the deformed Ricci tensor is not symmetric  will have a consequence on finding a consistent definition of the NC Einstein manifold. From this reason, in this paper we shall be concerned with a definition of the NC Einstein manifold that amounts not to a solution of the equation $Ric^{\star}(\partial_{\mu }, \partial_{\nu }) = 0$, but rather to a solution to its symmetrized form, which in the case of {\it{nice}}\footnote{See the next section for the explanation of the term {\it{nice}} basis.} basis corresponds to the symmetrized NC Ricci tensor
\begin{equation} \label{NCeinstein}
	Ric^{\star}_{(\mu \nu)} = \frac{1}{2} \Big( Ric^{\star}(\partial_{\mu }, \partial_{\nu }) + Ric^{\star}(\partial_{\nu }, \partial_{\mu }) \Big) = 0.
\end{equation}
In the case of deformation with a most general Drinfeld twist, the actual  NC Einstein equation has a more elaborate structure which incorporates $R$-symmetry. Its precise form can be found in \cite{glavniclanak}, as well as other details. However, for the special family of Drinfeld twists
that we consider here (starting from the next section, see (\ref{semi-Killingtwist})), it reduces to the special form (\ref{NCeinstein}).
In a situation that the deformation is carried by the same  special family of twists,  it might be tempting to contemplate about generalizing the Einstein tensor to a noncommutative case.
Namely, if we symmetrize the Ricci tensor as above and  define the Ricci scalar as $R^{\star} = g^{\star\mu \nu} \star Ric^{\star}_{(\mu \nu)}$, then
 postulating the vacuum Einstein equation in the form
\begin{equation}  \label{NCeinsteintensor}
	Ric^{\star}_{(\mu \nu)} - \frac{1}{2}g_{\mu \nu} \star R^{\star} = 0
\end{equation}
 straightforwardly  implies that $Ric^{\star}_{(\mu \nu)} =0,$ as can be clearly seen by $\star$-multiplying (\ref{NCeinsteintensor}) with  $g^{\star \rho \mu}$ 
from the left. While  this might be  an  indicator that defining the Einstein tensor as in (\ref{NCeinsteintensor}) proofs correct, it yet remains to see if this object is divergentless with respect to the $\star$-connection.


\section{Linearized noncommutative gravity on the Schwarzschild background}

To induce spacetime noncommutativity we will use the  twist of the form
\begin{equation}  \label{semi-Killingtwist}
	\ff = \exp {\Big(-i \frac{a}{2} \big(K \ot X - X \ot K \big)\Big)},
\end{equation}
where $X$ is some vector field and $K$ is a Killing field of the background metric. It is a Drinfeld twist, i.e. satisfies the  conditions (\ref{2cocyclecondition}).
Our Lie algebra of diffeomorphisms  is therefore two-dimensional, being generated by $X$ and $K$.
In addition, we require these two fields to commute, thus $\ff$ is an Abelian twist.
We will use the standard spherical coordinates $(t,r,\theta,\vp)$ and demand that basis consisting of their generators $(\dd_t, \dd_r, \dd_\theta, \dd_\vp)$ commutes with $X$ and $K$.
A basis of vector fields of this kind is referred to as a {\it{nice}} basis.
All of these properties are satisfied for 
\begin{equation} \label{kxdef}
K = \alpha \dd_t + \beta \dd_\varphi, \qquad X = \dd_r 
\end{equation}
 on the Schwarzschild background in spherical basis.
Because of the nice basis and Abelian twist, many formulas look like their commutative counterparts with $\star$-product replacing the usual pointwise product.
The $\star$-product according to (\ref{starproductdef}) is
\begin{equation*}
	f \star g = fg + i\frac{a}{2}\Big(K(f)X(g) - X(f)K(g)\Big) + O(a^2).
\end{equation*}
We use the twist   (\ref{semi-Killingtwist}) to study a noncommutative deformation of a linearized gravitational perturbation theory. The background metric $\oo{g}_{\mu \nu}$ is Schwarzschild:
\begin{equation*}
d s^2=-\left(1-\frac{2M}{r}\right)  d t^2+\frac{1}{1-  2M / r} d r^2+r^2\left(d \theta^2+\sin ^2 \theta d \vp^2\right), 
\end{equation*}
To study perturbations of the Schwarzschild spacetime we split the metric into background $\oo{g}$ and perturbation $h$,
\begin{equation}  \nonumber
    g_{\mu \nu} = {\mathring{g}}_{\mu \nu} + h_{\mu \nu},  \quad  \quad  {\mathring{g}}^{\mu \nu} {\mathring{g}}_{\nu \lambda}  = \delta^{\mu}_{~\lambda}.
\end{equation}
As already said, $K$ is the Killing vector field for the background (unperturbed) metric ${\mathring{g}}_{\mu \nu},$
\begin{equation}  \nonumber
    K({\mathring{g}}_{\mu \nu}) =  {\mathcal{L}}_K ({\mathring{g}}_{\mu \nu}) =0,  \quad   K({\mathring{g}}^{\mu \nu}) =  {\mathcal{L}}_K ({\mathring{g}}^{\mu \nu}) =0.
\end{equation}
The  metric $h$ is assumed to be small relative to $\oo{g}$ and thus we do our calculations up to the first order in $h$. When switching on a deformation, which is controlled by the parameter of deformation $a,$ we also keep only first order correction terms in $a.$ It will turn out that perturbation and deformation parts come in pairs, and are thus always coupled.

Due to relative simplicity of the twist (\ref{semi-Killingtwist})  and the nice basis, the conditions   (\ref{metricinversecondition})  for the metric inverse simplify significantly and
reduce to a rather familiar form, except only for the pointwise multiplication being replaced by a $\star$ multiplication:
\begin{equation}    \nonumber
   g^{\star \sigma \rho} \star g_{\rho \nu} = \delta^{\sigma}_{~\nu}  \quad \quad   g_{\nu \rho} \star  g^{\star  \rho \sigma} = \delta_{\nu}^{~\sigma}.
\end{equation}
Restricting to the  linear perturabations  $(\sim  O(h))$ and the  leading order in NC deformation $(\sim  O(a))$, the solutions to the  above stated conditions give for the inverse metric
\begin{equation}  \label{ginverse}
    g^{\star \mu \nu} = {\mathring{g}}^{\mu \nu} -  h^{\mu \nu} + {\tilde{g}}^{\mu \nu} = g^{\mu \nu} -  g^{\mu \rho} g_{\rho \lambda} \wedge  g^{\lambda \nu},
\end{equation}
where the following abbreviation was used
\begin{equation}
  f \wedge g = i\frac{a}{2} \big( K \otimes X - X \otimes K \big) (f \otimes g) = i\frac{a}{2} \bigg( K(f) X(g) - X(f)  K(g) \bigg).   \nonumber
\end{equation}
The inverse metric of $g_{\mu \nu}$ may also be written in the form
\begin{equation} 
	g^{\star \mu \nu} = \oo{g}^{\mu \nu} - \oo{g}^{\mu \alpha} \star h_{\alpha \beta} \star \oo{g}^{\beta \nu},  \nonumber
\end{equation}
which manifestly shows that a deformation, when present, is always coupled to  perturbation $h_{\mu \nu}.$
It turns out that such $\star$-inverse metric is not symmetric but Hermitian.

From the metric compatibility condition and relations (\ref{torsion-riemann-coef}) it then follows \cite{Aschieri:2009qh},\cite{schenkel} that 
coefficients of  the Levi-Civita  connection  in the nice basis, torsion, and Riemann curvature tensor are  in local coordinates given  by
\begin{align*}  \nonumber
	\Gamma^{\star \mu}_{~~\nu \rho} \equiv \Omega^{\mu}_{~\nu \rho} &= \frac{1}{2} g^{\star \mu \alpha} \star \big( \partial_{\nu} g_{\rho \alpha}  +  \partial_{\rho} g_{\nu \alpha} -  \partial_{\alpha} g_{\nu \rho} \big), \\
 T^{\star  \mu}_{~~~ \nu \rho} & = \Omega^{ \mu}_{~ \nu \rho} -  \Omega^{ \mu}_{~ \rho \nu}, \\
	R^{\star ~~~~\sigma}_{~\mu \nu \rho} &= \partial_{\mu} \Omega^{\sigma}_{~\nu \rho} - \partial_{\nu} \Omega^{\sigma}_{~\mu \rho} + \Omega^{\beta}_{~\nu \rho} \star \Omega^{\sigma}_{~\mu \beta} - \Omega^{\beta}_{~\mu \rho}  \star \Omega^{\sigma}_{~\nu \beta},
\end{align*}
which gives rise to the explicit form for the corrections to these quantities, all up to  the linear order in $h$ and $a$  $(\sim  O(h a)),$:
\begin{eqnarray}   \nonumber
    \Omega^{\mu}_{~ \nu \rho} &=& {\mathring{\Gamma}}^{ \mu}_{~\nu \rho} + \delta {\Gamma}^{ \mu}_{~ \nu \rho} + \big( {\mathring{g}}^{\mu \sigma} \wedge \delta {\Gamma}^{ \lambda}_{~ \nu \rho}  \big) {\mathring{g}}_{\sigma \lambda}  -  \big( h^{\mu \sigma} \wedge {\mathring{\Gamma}}^{ \lambda}_{~\nu \rho}  \big) {\mathring{g}}_{\sigma \lambda} \equiv \Gamma^{\mu}_{~ \nu \rho} +  {\tilde{\Omega}}^{\mu}_{~ \nu \rho},    \nonumber  \\
    T^{\star  \mu}_{~~~ \nu \rho} &=& \Omega^{ \mu}_{~ \nu \rho} -  \Omega^{ \mu}_{~ \rho \nu} = 0,   \nonumber \\
   R^{\star  ~~~~\sigma}_{~ \mu \nu \rho}  &=&  {\mathring{R}}^{ ~~~~\sigma}_{ \mu \nu \rho} + \delta  R^{~~~~ \sigma}_{\mu \nu \rho} + \partial_{\mu} {\tilde{\Omega}}^{\sigma}_{~\nu \rho} - \partial_{\nu} {\tilde{\Omega}}^{\sigma}_{~\mu \rho} + {\mathring{\Gamma}}^{ \beta}_{~\nu \rho} \wedge \delta {\Gamma}^{ \sigma}_{~ \mu \beta}
     +  \delta {\Gamma}^{ \beta}_{~ \nu \rho} \wedge {\mathring{\Gamma}}^{ \sigma}_{~\mu \beta}   \nonumber  \\    \nonumber
&+&    {\mathring{\Gamma}}^{ \beta}_{~\nu \rho} {\tilde{\Omega}}^{\sigma}_{~\mu \beta}
   + {\tilde{\Omega}}^{\beta}_{~\nu \rho} {\mathring{\Gamma}}^{ \sigma}_{~\mu \beta} -  {\mathring{\Gamma}}^{ \beta}_{~\mu \rho} \wedge   \delta {\Gamma}^{ \sigma}_{~ \nu \beta}  -   \delta {\Gamma}^{ \beta}_{~ \mu \rho} \wedge {\mathring{\Gamma}}^{ \sigma}_{~\nu \beta} - {\mathring{\Gamma}}^{ \beta}_{~\mu \rho}
   {\tilde{\Omega}}^{\sigma}_{~\nu \beta} - {\tilde{\Omega}}^{\beta}_{~\mu \rho}  {\mathring{\Gamma}}^{ \sigma}_{~\nu \beta},    \nonumber
\end{eqnarray}
and consequently leads to  the corrections within the same order in the vaccuum Einstein equations. Here $\delta \Gamma^{\mu}_{~~ \nu \rho}$ and $R_{\mu \nu \rho}^{~~~~\sigma}$ are just the $h$-linear parts without the noncommutativity corrections.
Taking  advantage of  the explicit form for the twist  (\ref{semi-Killingtwist}), and the above relation for  deformed Riemann tensor,
the NC Ricci tensor follows as
\begin{eqnarray}   \label{NCRicci}
 Ric^{\star }_{~ \nu \rho } &=& R^{\star  ~~~~\mu}_{~ \mu \nu \rho}  =  {\mathring{R}}^{ ~~~~\mu}_{ \mu \nu \rho} + \delta  R^{~~~~ \mu}_{ \mu \nu \rho}  
+ \partial_{\mu} {\tilde{\Omega}}^{\mu}_{~\nu \rho} - \partial_{\nu} {\tilde{\Omega}}^{\mu}_{~\mu \rho}  
 + i \frac{a}{2} \Big[  K( \delta {\Gamma}^{ \beta}_{~ \nu \rho}) X({\mathring{\Gamma}}^{ \mu}_{~\mu \beta})
   - X({\mathring{\Gamma}}^{ \beta}_{~\nu \rho}) K(\delta {\Gamma}^{ \mu}_{~ \mu \beta}) \Big] \nonumber  \\
&-& i \frac{a}{2} \Big[  K( \delta {\Gamma}^{ \beta}_{~ \mu \rho}) X({\mathring{\Gamma}}^{ \mu}_{~\nu \beta})
   - X({\mathring{\Gamma}}^{ \beta}_{~\mu \rho}) K(\delta {\Gamma}^{ \mu}_{~ \nu \beta}) \Big]  
+   {\mathring{\Gamma}}^{ \beta}_{~\nu \rho} {\tilde{\Omega}}^{\mu}_{~\mu \beta}
   + {\tilde{\Omega}}^{\beta}_{~\nu \rho} {\mathring{\Gamma}}^{ \mu}_{~\mu \beta}    - {\mathring{\Gamma}}^{ \beta}_{~\mu \rho}  {\tilde{\Omega}}^{\mu}_{~\nu \beta} - {\tilde{\Omega}}^{\beta}_{~\mu \rho}  {\mathring{\Gamma}}^{ \mu}_{~\nu \beta} \nonumber  \\
   &\equiv&   {\mathring{Ric}}_{ \nu \rho} +    \delta  Ric^{\star}_{ \nu \rho},  
\end{eqnarray}
where
\begin{equation}  \nonumber
  {\tilde{\Omega}}^{\mu}_{~\nu \rho} =  i \frac{a}{2} \Big[  K( - h^{\mu  \sigma}) X({\mathring{\Gamma}}^{ \lambda}_{~\nu \rho})
   - X({\mathring{g}}^{  \mu \sigma}) K(\delta {\Gamma}^{ \lambda}_{~ \nu \rho}) \Big] {\mathring{g}}_{\sigma \lambda}.
\end{equation}


Before writing down the equations describing NC gravitational perturbations, we  introduce the parameter $\lambda$ as an eigenvalue of the Killing field when acting   on the perturbation:
\begin{align*}
	h_{\mu \nu} \propto e^{-i \omega t}e^{i m \varphi} \implies \lie_K h_{\mu \nu} = i \lambda \: h_{\mu \nu}, \\ 
	\text{   for } \quad K = \alpha \dd_t + \beta \dd_\varphi, \quad
	\lambda = -\alpha \omega + \beta m.
\end{align*}


Note that    in case of the vanishing deformation, the equation  (\ref{NCeinstein}) reduces to the standard linear gravitational perturbation equation
\begin{equation}  \nonumber
	\delta Ric^{\star}_{\mu \nu} \xrightarrow {a \to 0} \delta Ric_{\mu \nu} = \nabla_{\rho} \: \delta \Gamma^{\rho}_{~\mu \nu} - \nabla_{\nu} \: \delta \Gamma^{\rho}_{~\mu \rho} = 0.
\end{equation}
The structure of the perturbation matrix under rotations on the $2$-sphere is given by 
\begin{equation}    \nonumber
  h_{\mu \nu} =
\left( \begin{array}{ccccc}
  S  & S  & V & V \\
   S   & S & V & V  \\ 
   V  & V & T &  T  \\
   V  & V & T & T  \\
\end{array} \right) .
\end{equation}
Since not all components transform like scalars, it is necessary to introduce the tensor spherical harmonics (generalized spherical harmonics) and expand perturbations $h_{\mu \nu}$ over them:

\begin{equation}  \nonumber
   h_{\mu \nu} (t,r,\theta, \varphi) = \sum_{\ell=0}^{\infty} \sum_{m=-\ell}^{\ell}  \sum_{n=1}^{10}  c_{n,\ell,m} (t,r) {\big( {\mathcal{Y}}^{(n)}_{\ell m} \big)}_{\mu \nu}.
\end{equation}
There are $10$ different types of them,

\begin{equation}   \nonumber
  {\mathcal{Y}}^{(1)}_{\ell m} = 
\left( \begin{array}{ccccc}
  1  & 0  & 0 & 0  \\
   0   & 0 & 0 & 0  \\ 
   0  & 0 & 0 &  0  \\
   0  & 0 & 0 & 0  \\
\end{array} \right)  Y_{\ell m},   \quad 
{\mathcal{Y}}^{(2)}_{\ell m} = \frac{i}{\sqrt{2}}
\left( \begin{array}{ccccc}
  0  & 1  & 0 & 0  \\
   1   & 0 & 0 & 0  \\ 
   0  & 0 & 0 &  0  \\
   0  & 0 & 0 & 0  \\
\end{array} \right)  Y_{\ell m}, \quad
  {\mathcal{Y}}^{(3)}_{\ell m} = 
\left( \begin{array}{ccccc}
  0  & 0  & 0 & 0  \\
   0   & 1 & 0 & 0  \\ 
   0  & 0 & 0 &  0  \\
   0  & 0 & 0 & 0  \\
\end{array} \right)  Y_{\ell m}, 
\end{equation}

\begin{equation}   \nonumber
{\mathcal{Y}}^{(4)}_{\ell m} = \frac{ir}{\sqrt{2\ell (\ell +1)}}
\left( \begin{array}{ccccc}
  0  & 0  & \partial_{\theta} Y_{\ell m} & \partial_{\varphi} Y_{\ell m}  \\
   0   & 0 & 0 & 0  \\ 
   \partial_{\theta} Y_{\ell m}  & 0 & 0 &  0  \\
   \partial_{\varphi} Y_{\ell m}  & 0 & 0 & 0  \\
\end{array} \right), \quad
{\mathcal{Y}}^{(5)}_{\ell m} = \frac{r}{\sqrt{2\ell (\ell +1)}}
\left( \begin{array}{ccccc}
  0  & 0  & 0 & 0  \\
   0   & 0 & \partial_{\theta} Y_{\ell m} &  \partial_{\varphi} Y_{\ell m}  \\ 
   0  & \partial_{\theta} Y_{\ell m} & 0 &  0  \\
   0  &  \partial_{\varphi} Y_{\ell m} & 0 & 0  \\
\end{array} \right),
\end{equation}

\begin{equation}   \nonumber
{\mathcal{Y}}^{(6)}_{\ell m} = \frac{r}{\sqrt{2\ell (\ell +1)}}
\left( \begin{array}{ccccc}
  0  & 0  & \frac{1}{\sin \theta} \partial_{\varphi} Y_{\ell m} & -\sin \theta \partial_{\theta} Y_{\ell m}  \\
   0   & 0 & 0 & 0  \\ 
   \frac{1}{\sin \theta} \partial_{\varphi} Y_{\ell m}  & 0 & 0 &  0  \\
   -\sin \theta \partial_{\theta} Y_{\ell m}  & 0 & 0 & 0  \\
\end{array} \right), \quad
 {\mathcal{Y}}^{(9)}_{\ell m} = \frac{r^2}{\sqrt{2}}
\left( \begin{array}{ccccc}
  1  & 0  & 0 & 0  \\
   0   & 0 & 0 & 0  \\ 
   0  & 0 & 1 &  0  \\
   0  & 0 & 0 & {\sin}^2 \theta  \\
\end{array} \right)  Y_{\ell m},  
\end{equation}

\begin{equation}   \nonumber
{\mathcal{Y}}^{(7)}_{\ell m} = \frac{ir}{\sqrt{2\ell (\ell +1)}}
\left( \begin{array}{ccccc}
  0  & 0  & 0 & 0  \\
   0   & 0 & \frac{1}{\sin \theta} \partial_{\varphi} Y_{\ell m} &  -\sin \theta \partial_{\theta} Y_{\ell m}  \\ 
   0  & \frac{1}{\sin \theta} \partial_{\varphi} Y_{\ell m} & 0 &  0  \\
   0  &  -\sin \theta \partial_{\theta} Y_{\ell m} & 0 & 0  \\
\end{array} \right),
\end{equation}

\begin{equation}   \nonumber
{\mathcal{Y}}^{(8)}_{\ell m} = \frac{-ir^2}{\sqrt{2\ell (\ell^2 -1)(\ell +2)}}
\left( \begin{array}{ccccc}
  0  & 0  & 0 & 0  \\
   0   & 0 &  0  &   0    \\ 
   0  & 0 & -\frac{2}{\sin \theta} \bigg( \partial_{\theta} \partial_{\varphi}  -  \cot \theta\partial_{\varphi} \bigg)  Y_{\ell m} &  \sin \theta \bigg(\partial^{2}_{\theta} - \cot \theta \partial_{\theta} - \frac{1}{\sin^2 \theta}    \partial^{2}_{\varphi} \bigg)Y_{\ell m} \\
   0  &  0 & \sin \theta \bigg(\partial^{2}_{\theta} - \cot \theta \partial_{\theta} - \frac{1}{\sin^2 \theta}    \partial^{2}_{\varphi} \bigg) Y_{\ell m} & 2 \sin \theta \bigg( \partial_{\theta} \partial_{\varphi}  -  \cot \theta \partial_{\varphi} \bigg) Y_{\ell m} \\
\end{array} \right),
\end{equation}

\begin{equation}   \nonumber
{\mathcal{Y}}^{(10)}_{\ell m} = \frac{r^2}{\sqrt{2\ell (\ell +1)(\ell -1)(\ell +2)}}
\left( \begin{array}{ccccc}
  0  & 0  & 0 & 0  \\
   0   & 0 &  0  &   0    \\ 
   0  & 0 &  \partial^{2}_{\theta} Y_{\ell m} &   \bigg(  \partial_{\theta} \partial_{\varphi}  -  \cot \theta \partial_{\varphi}   \bigg)Y_{\ell m} \\
   0  &  0 &    \bigg(  \partial_{\theta} \partial_{\varphi}  -  \cot \theta \partial_{\varphi}   \bigg)Y_{\ell m}     &  \bigg( \partial^2_{\varphi}  +  \sin \theta \cos \theta \partial_{\theta} \bigg) Y_{\ell m} \\
\end{array} \right),
\end{equation}
Tensor spherical harmonics behave differently under the parity transformation
\begin{equation}  \nonumber
  \bf{P}:  \bf{r} \longrightarrow  - \bf{r}   \quad
          [(\theta, \varphi)   \longrightarrow    (\pi - \theta,  \pi + \varphi) ].  \nonumber
\end{equation}
They all fall into  one of the two classes:  axial perturbations $[{(-1)}^{l+1}]$   or  polar  perturbations  $[{(-1)}^{l}]$. ~ ${\mathcal{Y}}^{(i)}, i=6,7,8$ are axial, the rest are polar.
In what follows we focus only on axial perturbations. Another important point is that there exists
 a freedom in choosing the local coordinates, which means that the change in coordinates  $x'^{\alpha} = x^{\alpha} - \xi^{\alpha} (x)$ amounts to gauging.
Under this gauge transformation, the perturbation tensor changes as
\begin{equation}  \nonumber
  h^{old}_{\mu \nu} \longrightarrow      h^{new}_{\mu \nu} =  h^{old}_{\mu \nu} + \nabla_{\mu}\xi_{\nu} + \nabla_{\nu}\xi_{\mu},  \nonumber
\end{equation}
with the appropriate gauge parameter $\xi^{\alpha}$. Here the covariant derivative is with respect to the background (Schwarzschild) connection. The choice of the gauge 
\begin{equation}  \nonumber
   \xi^{\mu} = \bigg( 0, 0, \Lambda(t,r) \frac{1}{\sin \theta} \partial_{\varphi} Y_{\ell m}, -\Lambda(t,r) \frac{1}{\sin \theta} \partial_{\theta} Y_{\ell m}  \bigg),  
\end{equation}
which is known as the Regge-Wheeler gauge  \cite{Regge:1957td}, significantly simplifies the form of the perturbation tensor
\begin{equation}    \nonumber
  h_{\mu \nu} = 
\left( \begin{array}{ccccc}
  0  & 0  &  -h_{0a} (t,r) \frac{1}{\sin \theta} \partial_{\varphi} Y_\ell^{~m} &  h_{0a} (t,r) \sin \theta \partial_{\theta} Y_\ell^{~m} \\
   0   & 0 & -h_{1a} (t,r) \frac{1}{\sin \theta} \partial_{\varphi} Y_\ell^{~m} & h_{1a} (t,r) \sin \theta \partial_{\theta}  Y_\ell^{~m}  \\ 
   -h_{0a} (t,r) \frac{1}{\sin \theta} \partial_{\varphi} Y_\ell^{~m}  & -h_{1a} (t,r) \frac{1}{\sin \theta} \partial_{\varphi} Y_\ell^{~m}  & 0 &  0  \\
    h_{0a} (t,r) \sin \theta \partial_{\theta} Y_\ell^{~m}  & h_{1a} (t,r) \sin \theta \partial_{\theta} Y_\ell^{~m} & 0 & 0  \\
\end{array} \right).
\end{equation}
Plugging this ansatz into  perturbation equation   (\ref{NCeinstein}), while parallelly exploiting  (\ref{NCRicci}),
leads to $10$ partial differential equations.   While $3$ among these $10$ equations are trivially satisfied, the remaining $7$  (after separating the angular and radial parts) may be broken down to a set of equations among which only $3$  have
distinctive radial parts:
\begin{eqnarray}  \nonumber
  Ric^{\star}_{(r\varphi)} &=&   \Bigg[  \frac{1}{1- \frac{2M}{r}} \bigg( \partial^2_t h_{1a} + \frac{2}{r} \partial_t h_{0a} - \partial_r \partial_t h_{0a} \bigg) +\frac{h_{1a}}{r^2} \Big( \ell (\ell +1) -2  \Big)      \nonumber   \\
&+&  \lambda a \bigg(  \frac{r - 4M}{r (r- 2M)^2} \partial_t h_{0a}  -  \frac{M}{ (r- 2M)^2} \partial^2_t h_{1a}   -  \frac{\ell(\ell + 1)+12}{ r^3} h_{1a}  + \frac{9M}{r^4} h_{1a}  
  \nonumber \\
&+&  \frac{M}{ (r- 2M)^2} \partial_r  \partial_t h_{0a}   - \frac{r - 2M}{ r^3} \partial_r h_{1a}  + 6 \frac{h_{1a}}{r^3}
 \bigg)  \Bigg]  \sin \theta \partial_{\theta} Y_{\ell m}(\theta, \varphi) =0,  \nonumber  \\
  Ric^{\star}_{(\theta \varphi)} &=&  \Bigg[ \partial_t h_{0a} - \frac{2M}{r^2} \Big(1- \frac{2M}{r} \Big) h_{1a} + \Big( \frac{4M}{r} -  \frac{4M^2}{r^2} -1  \Big) \partial_r h_{1a} 
  -\lambda a \Big(1 - \frac{2M}{r} \Big) \bigg( \frac{M }{ {(r - 2M)}^2} \partial_t h_{0a}    \nonumber \\
     &+&   \frac{3}{r^2} \bigg(1 - \frac{2M}{r} \bigg)  h_{1a} + \frac{M}{ r^2}   \partial_r h_{1a} \bigg)  \Bigg]
    \bigg( \sin \theta {\partial}^2_{\theta} - \cos \theta \partial_\theta - \frac{1}{\sin \theta} {\partial}^2_{\varphi } \bigg) Y_{\ell m}(\theta, \varphi) =0,
  \nonumber \\
 Ric^{\star}_{(t \varphi)} &=&  \Bigg[    \Big(1 - \frac{2M}{r} \Big)  \bigg( \frac{1}{r} \partial_t h_{1a}   + \frac{1}{2} \partial_r \partial_t h_{1a}  
    - \frac{1}{2} {\partial_r}^2  h_{0a}  \bigg) + \bigg( \ell (\ell + 1)  - \frac{4M}{r} \bigg)\frac{h_{0a}}{2 r^2}    \nonumber  \\
  &+ &    \frac{\lambda a}{4 r^4}  \bigg( -\Big(  2\ell(\ell + 1)  r + 2M \Big)h_{0a} +  2r  (2 r - 3M) \partial_t h_{1a} - 2r (2r - 5M) \partial_r h_{0a}  \nonumber  \\
 &+&  2M  r^2  \Big( \partial_r \partial_t h_{1a} -   \partial^2 _r h_{0a} \Big)  \bigg) \Bigg]  \sin \theta \partial_{\theta} Y_{\ell m}(\theta, \varphi)  = 0.   \nonumber
\end{eqnarray}

\begin{figure}[h] 
\vspace{-2ex}
\includegraphics[width=16cm]{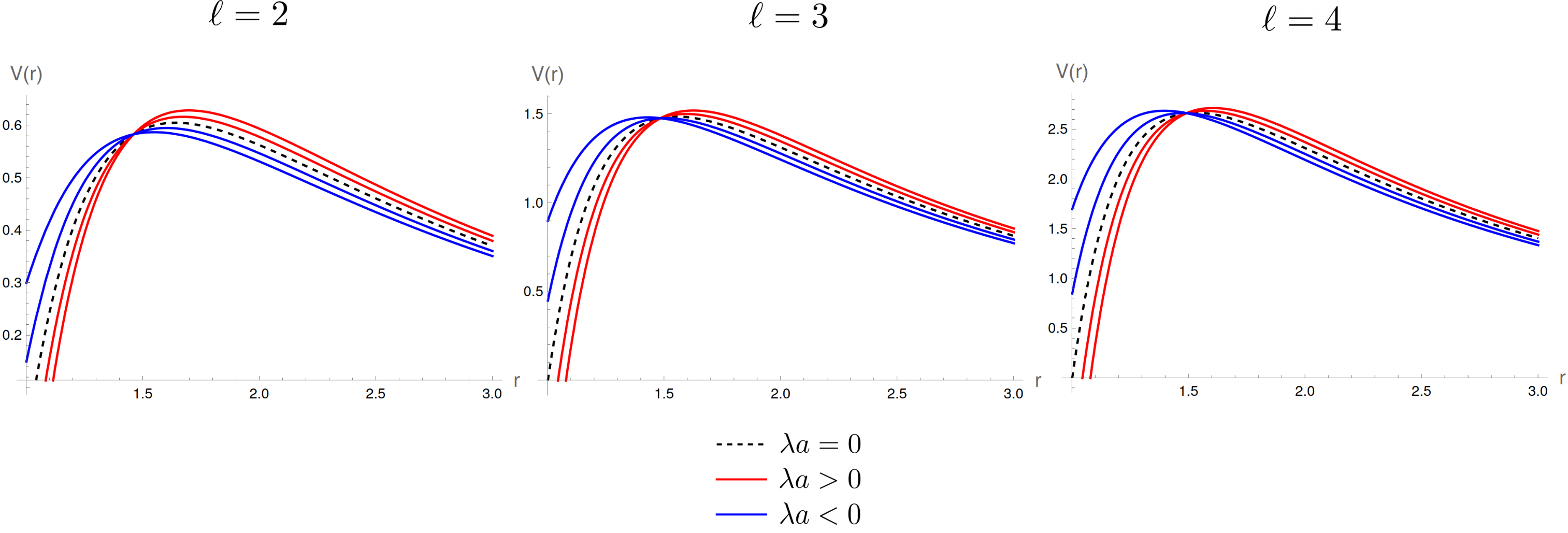}
	\caption{Plot of the potential with respect to the radial coordinate $r$ for $\ell = 2, 3$ and $4$. The blue lines correspond to $\lambda a = 0.1, \: 0.2$ and the red lines to $\lambda a = -0.1, \: -0.2$. Schwazschild radius is at $2M = 1$ and the dashed line is a potential without the noncommutativity corrections.} \label{figure}
\vspace{-5ex}
\end{figure}

The radial parts of these equations are not all independent. As shown in  \cite{glavniclanak}, only two of them are independent. Combining  first two radial equations and 
 substituting $h_{1a}(r) = \frac{r^2}{r - 2M}Q(r)$ leads to the second-order differential equation for $Q(r),$
\begin{align} 
     & \lambda a \Big( \frac{2 \ell (\ell + 1) r (r - 2M)^2 - 2(3r - 4M)(r - M)(r - 4M) - 2r^4 M \omega^2}{2 r^3 (r - 2M)^3}Q + \frac{(r - 4M)(3r - 4M)}{r^2(r - 2M)^2}\partial_r Q   \nonumber  \\
  &+ \frac{M}{r(r - 2M)} \partial^2_r Q \Big) 
    + \frac{(r - 2M)(6M - \ell (\ell + 1) r) + r^4 \omega^2}{r^2 (r - 2M)^2}Q + \frac{2M}{r(r - 2M)}  \partial_r Q +  \partial^2_r Q    \label{qeq}  = 0. 
\end{align}
In order to further reduce (\ref{qeq}) to a Schr\"odinger-type  equation, we introduce the modified tortoise coordinate $r_*$ and reshape the function  $Q(r)$  as
\begin{align*}
	\frac{d r}{d r_*} &= 1 - \frac{2M}{r} + \lambda a \frac{M}{ r^2} \implies r_* = r + 2M \log \frac{r - 2M}{2M} + \frac{\lambda a}{2} \frac{2M}{r - 2M}, \\
	Q(r) &= \Big( 1 + \frac{\lambda a}{2} \Big( \frac{3}{r} - \frac{1}{r - 2M} + \frac{1}{2M} \log \frac{r}{r - 2M} \Big) \Big)\: W(r).
\end{align*}
This  leads to crystallizing  the equation that keeps control over  NC gravitational perturbations,
\begin{align*}
	&\frac{d^2 W}{d r_*^2} + \Big( \omega^2 - V(r) \Big) W = 0, \\
	&V(r) = \frac{(r - 2M)\big(\ell (\ell + 1)r - 6M\big)}{r^4} + \lambda a \frac{\ell(\ell + 1)(6M - 2r)r + 2M(5r - 16M)}{2 r^5}.
\end{align*}
The first part of the potential is the  Regge-Wheeler potential  governing the  axial perturbations of the Schwarzschild black hole and the second part is the correction coming from the spacetime noncommutativity. 
Plot of the potential for $\ell = 2, 3, 4$ and several values of $\lambda a$ is shown in the Figure \ref{figure}. \newline

Potentials on Fig.\ref{figure} exhibit two interesting features, one being a Zeeman-like splitting, similar to that observed in case of a charged scalar field in the vicinity of Reissner-Nordstr\"{o}m black hole  \cite{Ciric:2017rnf},\cite{DimitrijevicCiric:2019hqq}
and the other being the  smearing of the horizon \cite{glavniclanak}. Smearing of the horizon is visible on the Fig.\ref{figure} as differing zeros of the potential (located at $2M - \lambda a / 2$), but also from the tortoise coordinate's dependence on $\lambda a$.

 It is also instructive to inspect the asymptotic behaviour of the solution near the horizon ($r \to 2M$) and at the spatial infinity ($r \to \infty$),
\begin{align*}
	&r \to 2M: \quad \frac{d^2 W}{d r_*^2} + \Big( \omega^2 + \lambda a \frac{3 - \ell(\ell + 1)}{16 M^3} \Big) W = 0 \implies W \propto e^{\pm i \Omega r_*}, \\
	&r \to \infty: \quad \frac{d^2 W}{d r_*^2} + \omega^2 W = 0 \implies W \propto e^{\pm i \omega r_*},
\end{align*}
where $\Omega^2 = \omega^2 + \lambda a \frac{3 - \ell(\ell + 1)}{16 M^3}$. 
The near horizon solution is modified with respect to the commutative theory, while the solution at infinity stays the same. This is expected to have an  impact on the quasinormal mode frequencies \cite{drugiclanak}.


\ack
A.S. would like to thank prof. \v{C}. Burdik and the organizers of the {\it XII. International Symposium on Quantum Theory and Symmetries (QTS12)} (Prague, July 2023) for their hospitality.
This  research was supported by the Croatian Science
Foundation Project No. IP-2020-02-9614 \textit{Search for Quantum spacetime in Black Hole QNM spectrum and Gamma Ray Bursts}.

\end{document}